\documentclass[pra,twocolumn,notitlepage,superscriptaddress,longbibliography]{revtex4-1}
\usepackage{amsmath}
\usepackage{amssymb}

\usepackage[unicode=true,colorlinks=true,citecolor=blue,urlcolor=blue]{hyperref}

\usepackage{bm}
\usepackage{color}
\usepackage{epsfig}

\usepackage[normalem]{ulem}

\begin{document}
\title{Radiative topological biphoton states in modulated qubit arrays}
	
\author{Yongguan Ke}
\affiliation{Guangdong Provincial Key Laboratory of Quantum Metrology and Sensing $\&$ School of Physics and Astronomy,Sun Yat-Sen University (Zhuhai Campus), Zhuhai 519082, China}
\affiliation{Nonlinear Physics Centre, Research School of Physics, Australian National University, Canberra ACT 2601, Australia}
	
\author{Janet Zhong}
\affiliation{Nonlinear Physics Centre, Research School of Physics, Australian National University, Canberra ACT 2601, Australia}	
	
\author{Alexander~V.~Poshakinskiy}
\affiliation{Ioffe Institute, St. Petersburg 194021, Russia}
	%\affiliation{Nonlinear Physics Centre, Canberra ACT 2601, Australia}
	
\author{Yuri~S.~Kivshar}
\affiliation{Nonlinear Physics Centre, Research School of Physics, Australian National University, Canberra ACT 2601, Australia}
\affiliation{ITMO University, St. Petersburg 197101, Russia}
	
\author{Alexander N.~Poddubny}
\email{poddubny@coherent.ioffe.ru}
\affiliation{Nonlinear Physics Centre, Research School of Physics, Australian National University, Canberra ACT 2601, Australia}
\affiliation{Ioffe Institute, St. Petersburg 194021, Russia}
\affiliation{ITMO University, St. Petersburg 197101, Russia}
	
\author{Chaohong Lee}
\email{lichaoh2@mail.sysu.edu.cn}
\affiliation{Guangdong Provincial Key Laboratory of Quantum Metrology and Sensing $\&$ School of Physics and Astronomy,Sun Yat-Sen University (Zhuhai Campus), Zhuhai 519082, China}
\affiliation{State Key Laboratory of Optoelectronic Materials and Technologies, Sun Yat-Sen University (Guangzhou Campus), Guangzhou 510275, China}

\begin{abstract}
We study topological properties of bound pairs of photons in spatially-modulated qubit arrays (arrays of two-level atoms) coupled to a waveguide. While bound pairs behave like Bloch waves, they are topologically nontrivial in the parameter space formed by the center-of-mass momentum and the modulation phase, where the latter plays the role of a synthetic dimension. In a superlattice where each unit cell contains three two-level atoms (qubits), we calculate the Chern numbers for the bound-state photon bands, which are found to be $(1,-2,1)$. For open boundary condition, we find exotic topological bound-pair edge states with radiative losses. Unlike the conventional case of the bulk-edge correspondence, these novel edge modes not only exist in gaps separating the bound-pair bands, but they also may merge with and penetrate into the bands. By joining two structures with different spatial modulations, we find long-lived interface states which may have applications in storage and quantum information processing.
\end{abstract}
	
\date{\today}
\maketitle

\section{Introduction}

Topological photonics has attracted a lot of attention recently~\cite{lu2014topological,Ozawa2019}.
Seminal works on topological photonics focused on  basic topological effects, such as topological edge states~\cite{wang2009observation,Hafezi2011,hafezi2013imaging}, Floquet topological insulators~\cite{rechtsman2013photonic,maczewsky2017observation}, Weyl and Dirac points~\cite{lu2015experimental,lu2016symmetry}. However, due to their unique optical properties, topological photonic structures can go beyond the straightforward generalisations of topological solid states systems.
In photonic systems, gain and loss can be easily engineered~\cite{suchkov2016nonlinear} and light-matter interaction can be tailored at will~\cite{Konstantin2016,Roy2017,KimbleRMP2018}. Therefore, there exist multiple non-Hermitian photonic systems which not only support novel topological states~\cite{lee2016anomalous,yao2018edge,gong2018topological,MartinezAlvarez2018}, but may also lead to novel optical devices and unusual applications~\cite{bandres2018topological,harari2018topological,klembt2018exciton,kruk2019nonlinear}.

Radiative topological edge states have already been predicted in linear optical systems~\cite{PhysRevLett.112.107403}.
Topological lasers and exciton-polariton topological insulators have also been demonstrated experimentally ~\cite{bandres2018topological,harari2018topological,klembt2018exciton,corzo2019waveguide}. However, topological photonics  is being explored  mainly
in the linear regime or classical-optics limit which do not take into account any effects of photon-photon interactions.
	
Strong photon-photon interactions are known to appear in photonic waveguides coupled to atoms or arrays of superconducting qubits~\cite{Konstantin2016,Roy2017,KimbleRMP2018}. These quantum waveguides support various exotic correlated states such as photon bound states~\cite{Zhang2019arXiv}, novel twilight states~\cite{Ke2019}, and self-induced localized states~\cite{zhong2019photonmediated}, which look very promising for storage and processing of quantum information. 
By arranging the positions of qubits and designing the waveguide structure, the topological edge states have been analyzed in single-excitation systems~\cite{PhysRevLett.119.023603,bello2019unconventional,Dowing2019,Yu12743}.
However, exotic effects may emerge in interacting topological systems when two or more particles (or quasi-particle excitations) interact~\cite{ANP2017,ANP2017b,KePRA2017,Qin2017,Qin_2018,lin2019interaction}. Being analytically and numerically more challenging, the study of an interplay between photon interaction, non-Hermiticity, and topology is an exciting novel avenue to be explored in this field.

\begin{figure}[!b]
	\includegraphics[width=0.45\textwidth]{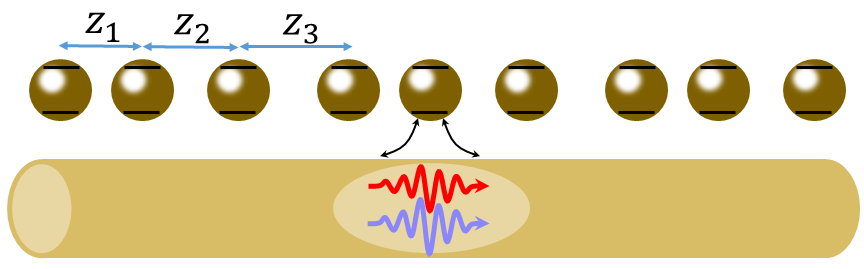}
	\caption{Schematic of a spatially modulated array of qubits coupled to a waveguide. The modulation period contains 3 qubits. Coupling between different
qubits is mediated by photons propagating along the waveguide.
	}\label{Diagram}
\end{figure}

In this paper, we study the topological properties of two excitations in an atomic array with spatial modulation and photon-mediated long-range coupling. Due to the long-range hopping, even though double excitations are forbidden in an individual qubit, there exist two-excitation bound states whose probability amplitudes decay exponentially  with the distance between them.
By considering an infinite array with unit cell of $3$ two-level atoms (3 qubits), we concentrate on the bound-state bands.
By varying the phase of spatial modulation and the center-of-mass momentum of two quasi-particle excitations, we observe the emergence of three bound-state bands of nontrivial topological invariants characterized by a set of the Chern numbers $(-1,2,-1)$.
For a finite array, we find the radiative two-excitation topological edge state as a subradiant state with decay rate less than that of an individual qubit. In contrast to the conventional bulk-edge correspondence, these topological bound edge modes not only exist in the band gaps but also merge and penetrate into the bound-state bands. There also exist long-lived interface states between the  two arrays with different spatial modulation phases.
	
The paper is organized as follows. In Sec.~\ref{SecI}, we introduce the model describing excitations in our qubit array.
Next, we analyze the band structure and calculate the Chern numbers for the two-excitation bound-state bands (Sec.~\ref{SecII}).
In Sec.~\ref{SecIII}, we calculate the spectral and radiative topological states for a finite structure.
In Sec.~\ref{SecIV}, we analyze long-lived interface states in a system combining two qubit arrays with different spatial modulations.
Finally, Sec.~\ref{SecV} concludes the paper with a brief summary and discussion.

\section{Model}
\label{SecI}

We consider a spatially-modulated array of qubits (two-level atoms) coupled to a waveguide, see Fig.~\ref{Diagram}.
Similar to earlier studies~\cite{Ke2019,zhong2019photonmediated}, we describe the system by the Hamiltonian
\begin{equation}\label{Ham}
\hat H = -i\Gamma_0 \sum\limits_{j,l}^{} {b_j^\dag b_l^{\vphantom{\dag}}{e^{i\varphi|{z_j} - {z_l}|}}}  + \frac{\chi}{2} \sum\limits_{j}^{} {b_j^\dag b_j^\dag b_j^{\vphantom{\dag}}b_j^{\vphantom{\dag}}},
\end{equation}
where $\Gamma_0$ is the strength of a radiative decay for a single qubit,
$b_j$ ($b_j^\dagger$) are the annihilation (creation) operators for the bosonic excitations in the $j$-th qubit,
$z_j$ is the position of the $j$th qubit, and $\chi$ describes the on-site interaction.
The hopping of an excitation from $j$-th to $l$-th qubits is mediated by the waveguide photon which contributes a phase factor 
depending on the hopping distance. The phase constant $\varphi=\omega_0 d/c$ depends on the qubit resonant frequency $\omega_0$, 
the light velocity $c$, and the spacing $d$. The qubit positions are modulated according to the relation, 
\begin{equation}
z_j=j+\delta \cos(2\pi j /\beta+\phi_0),
\end{equation}
where $\beta$ is the spatial period, and $\phi_0$ is the modulation phase. We choose $\beta=3$, which means that we consider $3$ qubits per unit cell.
Such modulation, inspired by the Aubry-Andr\'e-Harper model~\cite{aubrey}, is known to give rise to not trivial topology of single-photon bands~\cite{Lang2012} and radiative edge states~\cite{PhysRevLett.112.107403}.

In the limit $\chi\rightarrow \infty$, double excitations in a single qubit are forbidden.
Due to the conservation of the excitation number, the Hilbert subspaces for different excitation numbers are decoupled.
Below, we only consider the subspace of two excitations, where the state can be expanded in the two-excitation basis as $|\psi\rangle=\sum_{j<l}\psi_{j,l}b_j^{\dagger}b_l^{\dagger}|0\rangle$.
For simplicity, we denote the basis $b_j^{\dagger}b_l^{\dagger}|0\rangle\equiv|j;l\rangle$.
Because the qubit array has the period $3$, we can also refer the $(j=3m+n)$-th qubit with two indices $(m,n)$, where $m$ is the cell index, and $n=1,2,3$ indicates the location in the unit cell.

\section{Band structure and the Chern number}
\label{SecII}

\begin{figure}[!t]
\includegraphics[width=0.5\textwidth]{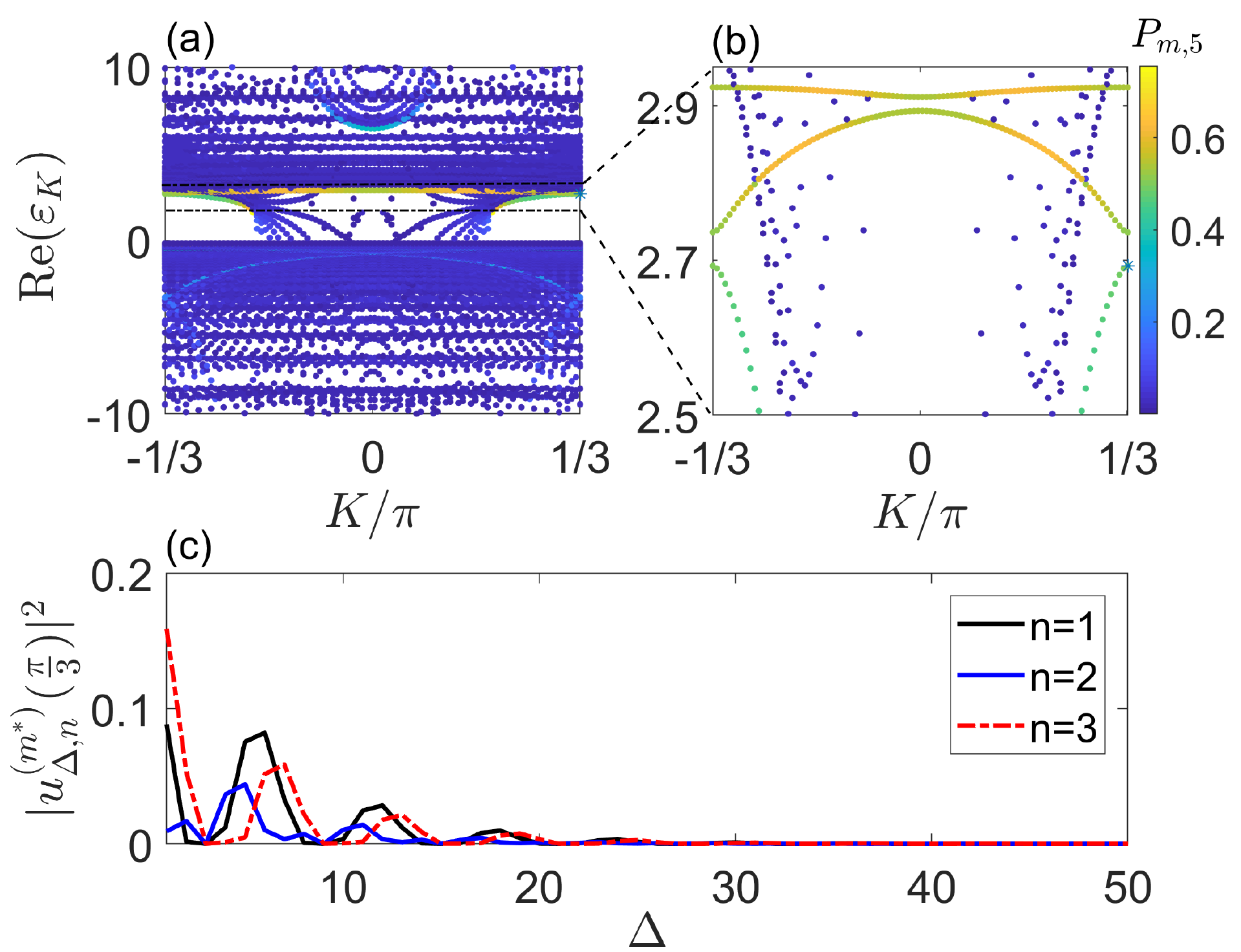}
\caption{(a) The band structure versus the center-of-mass momentum for two excitations. The colors denote the probability to find the two excitations within a certain short distance $\Delta_0=5$. (b) The zoom-in of (a) showing  the position of considered three bound-state bands. (c) The probability as a function of the distance $\Delta$ between two excitations. The parameters are chosen as $\Gamma_0=1$, $\delta=0.1$, $\varphi=0.3$, $L=99$ and $\phi_0=0$. }\label{ComBand}
\end{figure}

For an infinite structure, the total energy is invariant if spatial positions of the two excitations are shifted by a unit cell as a whole.
It means that the center-of-mass momentum is a good quantum number according to the many-body Bloch theorem~\cite{KePRA2017,Qin2017,Qin_2018,lin2019interaction}.
To present the Hamiltonian in a block-diagonal form with different momenta, we introduce a new basis
\begin{equation}\label{Basis}
|K,\Delta ,n\rangle  = \sum\limits_{m=1}^{L} {{e^{iK({z_{(m,n)}} + {z_{(m,n+\Delta) }})/2}}|m,n; m,n+\Delta \rangle }.
\end{equation}
Here, $K$ is the center-of-mass momentum, $L$ is the truncation number, $\Delta$ is the index difference between the two excitations.
Two states may be different even if they have the same value of $\Delta$.
Thus, we need to distinguish such different states by an additional index, $n$, the location of the first excitation in a unit cell.
In the new basis, the Hamiltonian elements are given by
\begin{equation}\label{HamBloch}
\langle K',\Delta ',n'|\hat H|K;\Delta ,n\rangle = - i{\Gamma _0}\delta_{K,K'}  M_{\Delta',n';\Delta,n},
\end{equation}
with
\begin{equation}
\begin{array}{l}
M_{\Delta',n';\Delta,n}= {e^{i\frac{K}{2}({z_{n + \Delta }} - {z_{n + \Delta '}})}}{e^{i\varphi |{z_{n + \Delta }} - {z_{n + \Delta '}}|}} f_{n - n'}  \\
+ {e^{i\frac{K}{2}({z_n} - {z_{n + \Delta  + \Delta '}})}}{e^{i\varphi |{z_n} - {z_{n + \Delta  + \Delta '}}|}}f_{n + \Delta  - n'} \\
+{e^{i\frac{K}{2}({z_{n + \Delta }} - {z_{n - \Delta '}})}}{e^{i\varphi |{z_{n + \Delta }} - {z_{n - \Delta '}}|}}f_{n' + \Delta ' - n} \\
+ {e^{i\frac{K}{2}({z_n} - {z_{n + \Delta  - \Delta '}})}}{e^{i\varphi |{z_n} - {z_{n + \Delta  - \Delta '}}|}}f_{n + \Delta  - \Delta ' - n'}
\end{array}
\end{equation}
where
\begin{equation}
f_{x}=\left\{ {\begin{array}{*{20}{c}}
	{1,\ \textrm{if} \mod(x,3) = 0;}\\
	{0,\ \textrm{if} \mod(x,3) \ne 0.}
	\end{array}} \right.
\end{equation}
The Hamiltonian elements are non-zero only when $K=K'$, in other words, there is no coupling between subspaces with different center-of-mass momenta. By solving the eigenvalue problem $-i\Gamma_0 \hat M |u^{(m)}(K)\rangle=2 E_{K,m}|u^{(m)}(K)\rangle$, we can obtain the energy bands depending on the center-of-mass momentum, see Fig.~\ref{ComBand}(a) and its zoom-in (b). Since the hopping terms are non-Hermitian, the eigenvalues are usually complex. However, the imaginary part of the energy vanishes as the number of qubits increases, and all the energies become real for an infinite array~\cite{PhysRevLett.112.107403}. For the truncation number $L=99$, the imaginary part of energy is quite small.

To characterize the bound states, we calculate the probability of finding two excitations within a certain short distance $\Delta_0$,
\begin{equation}\label{key}
{P_{m,\Delta_0}}(K) = \sum\limits_{1 \le n\le n_0;1\le \Delta  \le {\Delta _0}}^{} {{{\left| {u _{\Delta ,n}^{(m)}(K)} \right|}^2}}.
\end{equation}
Here, ${u _{\Delta,n}^{(m)}(K)}$ is the amplitude of the $m$-th eigenstate $|u^{(m)}(K)\rangle$.
The colors in Figs.~\ref{ComBand}(a) and (b) indicate the bounded probability $P_{m,\Delta_0}$.
Here, $n_0=3$ is the spatial modulation period, and the other parameters are chosen as $\Gamma_0=1$, $\delta=0.1$, $\varphi=0.3$, $\phi_0=0$ and $\Delta_0=5$ (actually $\Delta_0$ can be chosen as another value $\lesssim 10$ which does not affect the features of results).
We find that there exist three bound-state bands, merging into the scattering bands.
The origin of these three bound-state bands is similar to the formation of conventional energy bands.
For a simple periodic array, a bound-state band is folded in the reduced Brillouin zone ($-\pi/3 \le K\le \pi/3$).
The gaps between bands are open at the degenerate points as the modulation strength is increased from $0$.

We show the probability amplitude $| {u _{\Delta ,n}^{(m^*)}(K)}|^2$ of the bound state with $K=\pi/3$ as a function of the distance between two excitations, see Fig.~\ref{ComBand}(c).
$n=1,2,3$ indicate the position of the first qubit in a unit cell.
The corresponding energy is marked as $*$ in Fig.~\ref{ComBand}(a).
It is clear that the bound state exponentially decays with the relative distance $\Delta$, independent of the position of the qubit in a unit cell.
Apart from the bound states, there are many scattering states, which can be viewed as two independent free excitations.
The wavefunction of scattering state can be decomposed as the product of two single-excitation Bloch functions.
%\red{(Check)}
%\begin{equation}\label{Scattering}
%\psi_{j_1,j_2}(K)=\frac{1}{\sqrt {2}}\left[\xi_{j_1}\big(\frac{K+\kappa}{2}\big)\xi_{j_2}\big(\frac{K-\kappa}{2}\big)+(j_1\leftrightarrow j_2)\right], \nonumber
%\end{equation}
%Here, $\xi_{j}(k)$ is the amplitude at $j$th qubit for the single-excitation Bloch state $|\xi (k)\rangle$.
%$\kappa$ is the relative momemtum between the two exictations.
If we denote the eigenenergy of the single-excitation Bloch state as $\varepsilon_k$, then the energy for the scattering state is approximately given by
\begin{equation}\label{ScatEne}
E_K\approx \frac{\varepsilon_{(K+\kappa)/2}+\varepsilon_{(K-\kappa)/2}}{2}.
\end{equation}

\begin{figure}[!t]
\includegraphics[width=0.48\textwidth]{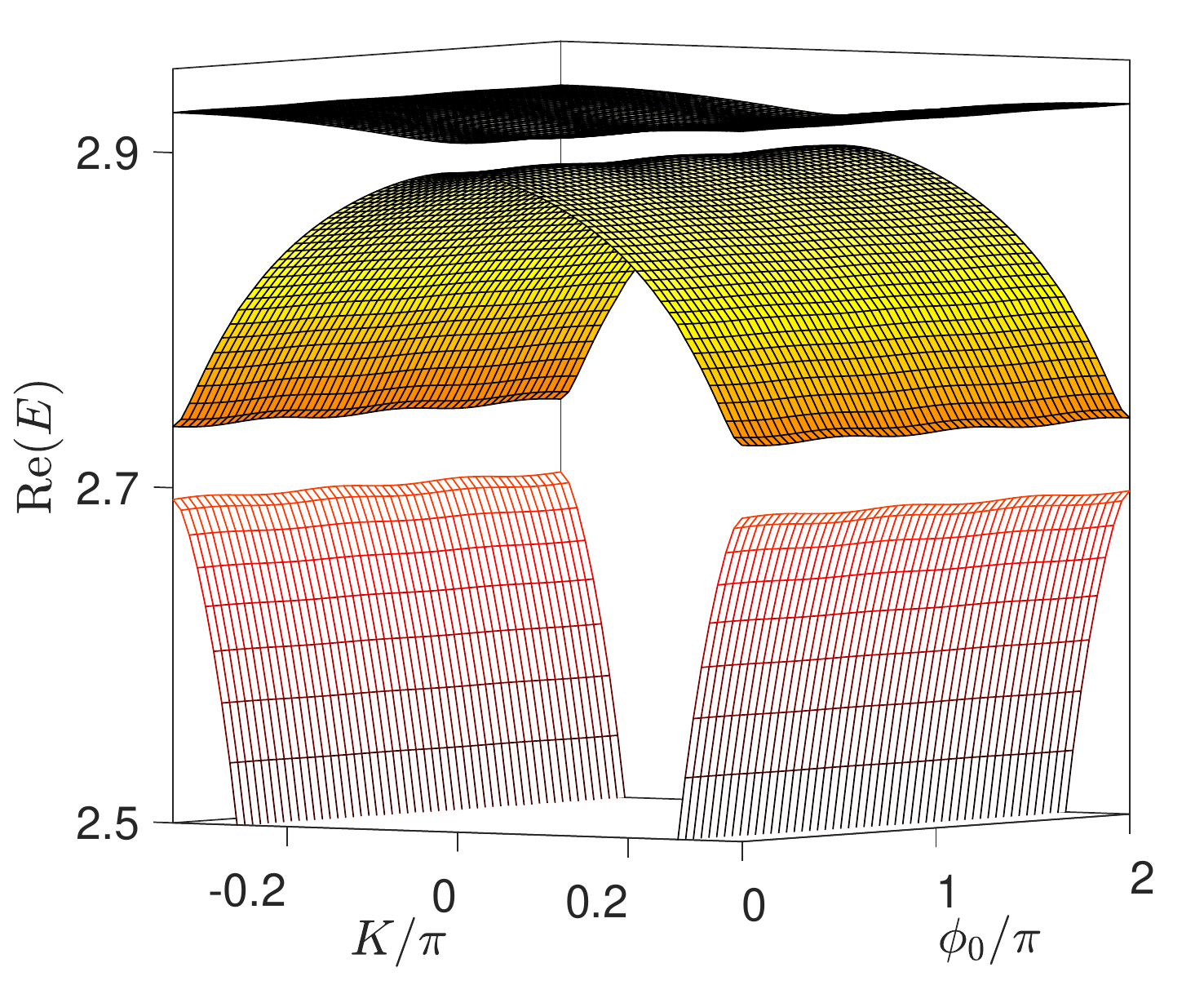}
\caption{The bound-state bands versus the center-of-mass momentum $K$ and the modulation phase $\phi_0$. The eigenenergy $E$ is in the unit of $\Gamma_0$. The parameters are chosen as $\delta=0.1\Gamma_0$, $L=70$ and $\varphi=0.3$}\label{BoundBand}
\end{figure}

Chern number is a well-known topological invariant which characterizes the band topology in the two-dimensional parameter space.
To reveal the topological nature of the bound-state bands, we calculate the Chern number for the bound-state bands in the $(K,\phi_0)$-plane.
The Chern number is defined as
\begin{equation}\label{Chern}
C_m=\frac{1}{2\pi}\int_{-\pi/3}^{\pi/3}dK\int_{0}^{2\pi}d\phi_0 \mathcal F_m(K,\phi_0),
\end{equation}
where the Berry curvature is $\mathcal F_m=i\big(\langle\partial_{\phi_{0}}u^{(m)}|\partial_{K}u^{(m)}\rangle -\langle\partial_{K}u^{(m)}|\partial_{\phi_{0}}u^{(m)}\rangle\big)$.
Because the energies of the bound-state bands merge into those of the scattering-state bands, we need to distinguish bound-state bands from the scattering-state ones.
Here, we pick out the bound-state bands if the bounded probability $P_{m,5}$ is greater than a threshold, $P_{m,5}>P_{th}$. The threshold value 
$P_{th}=0.25$ is selected by numerical experiments so that it can  distinguish the smooth and continuous surface of bound-state bands.
The bound-state bands depending on $K$ and $\phi_0$ are shown in Fig.~\ref{BoundBand}.
The top two bands are continuous in the $(K,\phi_0)$-plane, thus we can directly apply Eq.~\eqref{Chern} to calculate the Chern numbers. On the other hand,  the lowest band has two disconnected two branches, which have linear dispersion when $K$ approaches to $\pm K_0\approx \pm 0.19\pi$.
However, the major contributions for Chern number come from the Berry curvature around the band edge $K=\pm \pi/3$.
Thus, we sum the Berry curvature of the two branches to get the Chern number.
The Chern numbers for the three bands from bottom to the top are $(-0.9318,\ 1.9505,\ -0.9894)$ for the truncation number $L=70$, which tend to the ideal integers $(-1,\ 2,\ -1)$ as $L$ increases.
Similarly, we can also calculate the Chern numbers for scattering states, which originate from the single-particle topology.

\section{Radiative topological states}	
\label{SecIII}

\begin{figure*}[!htp]
	\includegraphics[width=0.94\textwidth]{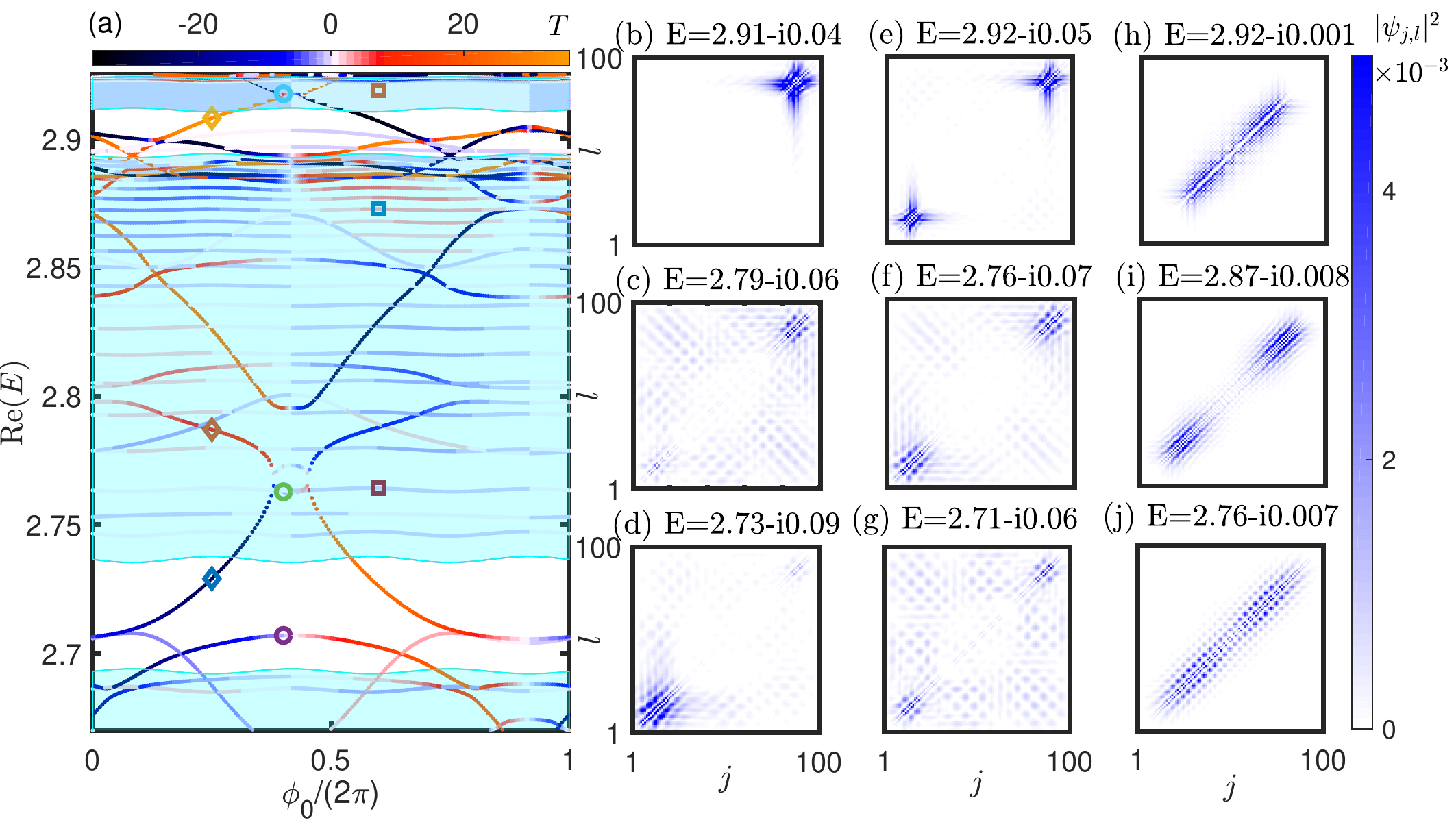}
	\caption{Energy spectrum and topological states in a finite structure. (a) Energy spectrum vs the modulation phase. The colors in (a) denote the tilted degree. (b)-(d) Eigenstates at $\phi_0=2\pi/5$, marked by diamonds from top to  bottom in (a).  (e)-(f) Eigenstates at $\phi_0=4\pi/5$, marked by circles from top to bottom in (a). (h)-(j) Eigenstates at $\phi_0=6\pi/5$, marked by squares from top to  bottom in (a). The colors in (b)-(j) denote the probability $|\psi_{j,l}|^2$.   }\label{TopoState}
\end{figure*}

In this section, we proceed to the spectrum and eigenstates in a finite and modulated array. There is always radiative loss for a finite array with qubit-photon coupling, that is, the excitations will be transferred into photons and escape from the array into the waveguide. The radiative decay rate is determined by the imaginary part of the eigenvalue, $ \Gamma=-{\rm Im\,}(\varepsilon)$~\cite{Ke2019}.
Thus, all of the eigenstates in a finite array have radiative losses.

We now study the energy spectrum as a function of the modulation phase, see Fig.~\ref{TopoState}.	
The parameters are chosen as $\Gamma_0=1$, $\varphi=0.3$, $\delta=0.1$ and the number of qubits is $N=100$.
We focus on the energy spectrum around the bound-state bands which are shaded in cyan, and check the bulk-edge correspondence.
To distinguish the edge and scattering states, we calculate the tilted degree,
\begin{equation}\label{Degree}
T=\sum\limits_{j,l} g [j+l-(N+1)]|\psi_{j,l}|^2,
\end{equation}
where $g$ is the tilted factor, which is set to $g=0.5$ in our calculation.
For $N=100$, $T$ takes the maximum and minimum values $\pm (N/2-2)=\pm 48$ if the two excitations are located at the leftmost and the rightmost edges, respectively. On the other hand, $T$ is close to zero  if the two excitations are in a scattering state or in an equal superposition of left and right edge states.
It means that $T$ can be used to select the states localized at one edge.
The colors in Fig.~\ref{TopoState}(a) denote the tilted degree of the corresponding eigenstates where black/dark blue and red signify left and right localized edge states respectively.
In the band gaps, we can find black and red curves entering the bound-state bands.
There also exists dark blue and red curves entirely merging in the bound states bands.

To study the correlation properties, we calculate the joint probability distribution, $|\psi_{j,l}|^2$, for one excitation at the $j$th qubit and the other one at the $l$th qubit.
For $\phi_0/(2\pi)=1/5$, we choose three typical states marked by diamonds from bottom to the top in Fig.~\ref{TopoState}(a), with the larger absolute values of the tilted degree.
The top diamond in the second gap represents the state when two excitations are bounded within short distance and close to the right edge, see Fig.~\ref{TopoState}(b).
It also has the crossing feature, which may be affected by the self-localized states with close energies~\cite{zhong2019photonmediated}.
Such state can be termed as bound-edge state.
The bottom diamond represents a bound-edge state localized at the left edge, see Fig.~\ref{TopoState}(d).
The middle diamond represents the bound-edge state slightly mixed with the scattering states, see Fig.~\ref{TopoState}(c).
This coexistence of bound-edge and scattering states in a single eigenstate confirms again that the bound-edge states not only exist in the band gaps, but also merge and penetrate into the bound state bands.

Next, we choose  we choose three states for the phase $\phi_0/(2\pi)$ close to $2/5$,  marked by circles from top to the bottom, which connect the dark blue and red curves, see Fig.~\ref{TopoState}(e)-(g).
These states are in the superposition of bound states localized at the left and right edges.
These symmetric or asymmetric bound-edge states have a decay rate in the order of $\varphi^2 \Gamma_0$.
It is interesting that the phase values $\phi_0/(2\pi)=2/5+1/60$ and $\phi_0/(2\pi)=2/5+1/60+1/2$ correspond to the critical points where all the states are symmetrically distributed with respect to the array center.
This is because the system preserves inversion symmetry at these critical points, that is, the energy is unchanged when changing the position of two excitations at $(j,l)$ to $(N+1-l,N+1-j)$.
The critical point is obtained by solving $\cos [2\pi /3(j + 1) + {\phi _0}] - \cos (2\pi /3j + {\phi _0}) = \cos [2\pi /3(N + 1 - j) + {\phi _0}] - \cos [2\pi /3(N - j) + {\phi _0}]$, that is, $\cos ({\phi _0} - 2\pi /3) + \cos ({\phi _0}) = 2\cos ({\phi _0} + 2\pi /3)$ for $N=100$. The critical points shift by multiples of $2\pi/3$ as the number of qubits changes.
Away from such critical points, the two-excitation states  tend to either the  left or the  right part of the array.

For $\phi_0/(2\pi)=3/5$, we focus on the three scattering states marked as squares in the bulk bands, see Fig.~\ref{TopoState}(h)-(j). The probability distributions are concentrated along the diagonal line $l=j$, which is a clear signature  of the bound states. The bound states in the bulk have a decay rate smaller that of the bound-edge states by an order of magnitude.
The intuitive explanation is that the radiative loss happens at the edge qubits and the bound states in the bulk have less occupation at the edge than that of the bound edge states~\cite{alex2019quasiflat}.
These bound states can be classified as subradiant state which have the radiative lifetime larger than that of the single qubit.

\begin{figure}[!t]
	\includegraphics[width=0.46\textwidth]{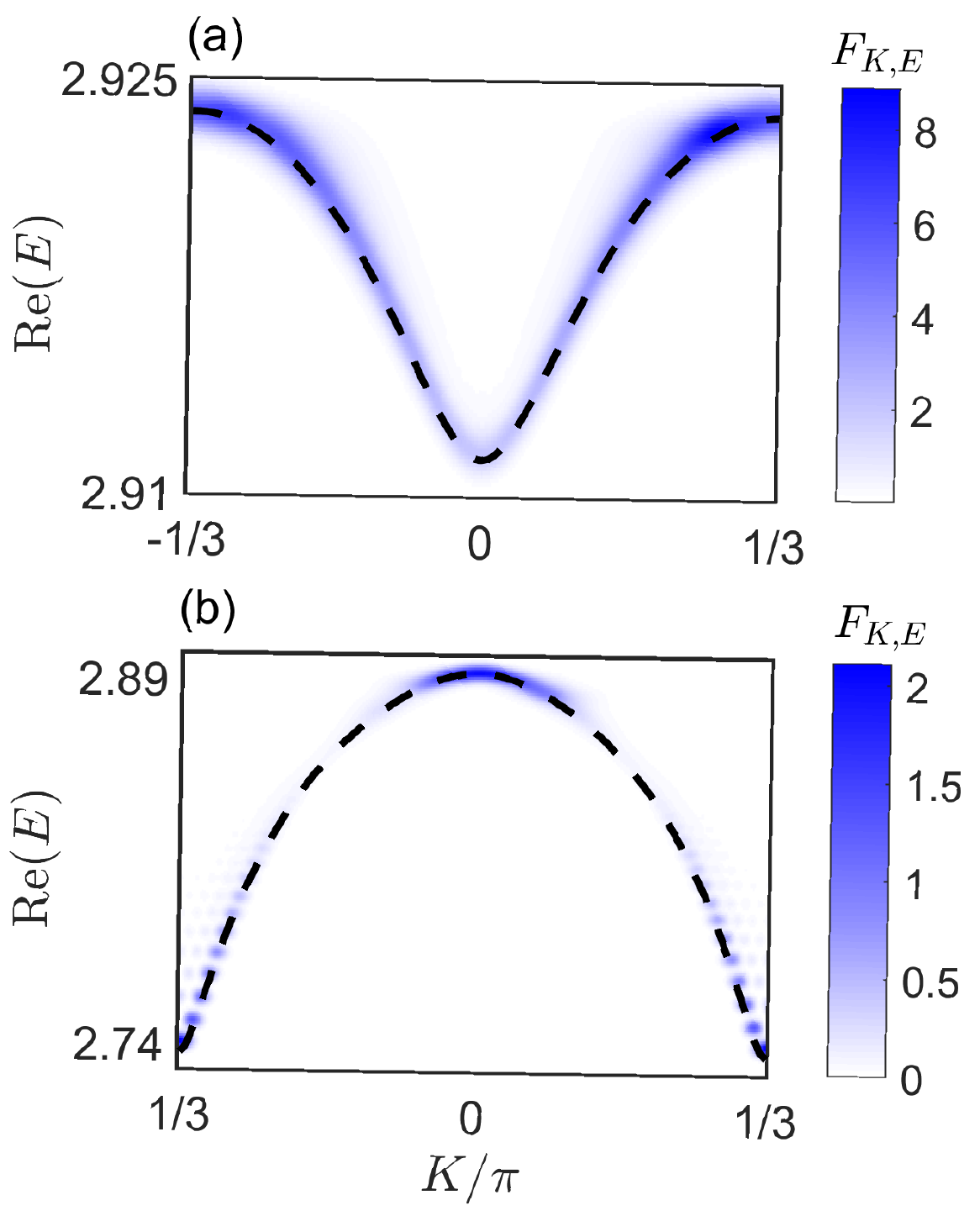}
	\caption{ Density distribution of bound states in the $(K, E)$-plane around (a) the third bound-state band and (b) the second bound-state band. The dashed lines indicate the third and second bound-state bands. The parameters are chosen the same as those in Fig.~\ref{ComBand}.   }\label{DenS}
\end{figure}

The center-of-mass motion of the bound states behaves as a Bloch wave with momentum $K$.
To extract the center-of-mass momentum, we make a Fourier transformation of the bound states:
\begin{equation}\label{Fourier}
{{\Psi }_{s,n}^{(v)}}(K) = \sum\limits_{m = 0}^{M-1} {\psi _{3m + n - s, 3m + n}^{(v)}{e^{ iK({z_{3m + n}} + {z_{3m + n - s}})/2}}},
\end{equation}
where $v$ refers to the $v$th eigenstates, $s$ restricts the distance between two excitations, and $n$ is the position index in a unit cell.
Since we know the center-of-mass momentum and the eigenvalues, we can reconstruct the dispersion relation by calculating the 
wave-vector-resolved density of states~\cite{Lyubarov2018},
\begin{equation}\label{DenOfState}
F(E,K) = \sum\limits_v^{} {{e^{ - \frac{{|E - {\rm Re}\: ({E_v}){|^2}}}{{2{\sigma ^2}}}}}\sum\limits_{s,n}^{} {{{\left| {\Psi _{s,n}^{(v)}(K)} \right|}^2}} },
\end{equation}
where the Gaussian broadening $\sigma$ is introduced for better visualization.
Figures~\ref{DenS}(a) and (b) show the density of states for the bound states in the second and third bound-state bands, respectively.
The parameters are chosen as $\Gamma_0=1$, $\varphi=0.3$, $\delta=0.1$, $\phi_0=0$,	$s=10$, $M=50$ and $\sigma=1\times 10 ^{-3}$. The value of $\sigma$ can be chosen differently, since it only affects the width of density distribution.
The dashed lines in Figs.~\ref{DenS}(a) and (b) show the second and third bound-state center-of-mass dispersion, respectively, and well match the density of states~Eq.~\eqref{DenOfState}. This means that our calculations in the finite structure and in the infinite structure are consistent.

%%%%%%%%%%%%%%%%%%%%%%%%%%%%%%%%%%%%

\section{Long-lived interface states}
\label{SecIV}	

\begin{figure}[!t]
\includegraphics[width=0.46\textwidth]{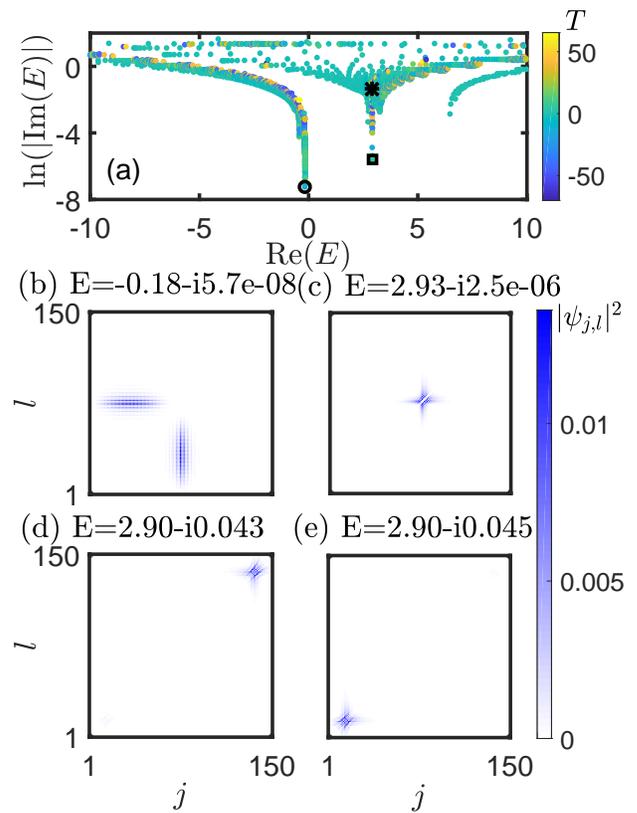}
\caption{ (a) The eigenvalues distribution in the complex plane. The colors denote tilted degree of eigenstates. (b)-(e) The probability distribution for two excited states with energies marked by $\circ$, $\square$, $\times$ and $+$ in (a), respectively. }\label{InterfaceS}
\end{figure}

According to our calculations,  the bound-edge states have the finite radiative decay rate on the order of $\sim \varphi^2\Gamma_0$.
This is because these states appear at an interface between the qubit array and free space, where the radiative loss becomes significant~\cite{alex2019quasiflat}. However, if we connect two qubit arrays with different modulation phases, there should be edge states with longer lifetime at the interface located in the bulk of array. The qubit at the interface plays the role of a defect which can trap the excitations for long time. To verify this argument, the positions of qubits are arranged as
\begin{equation}\label{Position}
{z_j} = \left\{ {\begin{array}{*{20}{c}}
	{j + \delta \cos (2\pi j /3),\;j \le N/2,}\\
	{j + \delta \cos (2\pi j/3 + \pi ),\; j > N/2.}
	\end{array}} \right.
\end{equation}
There is an interface between $N/2$ and $N/2+1$ qubits.

We calculate the energy distribution in the complex plane, as shown in Fig.~\ref{InterfaceS}
for the parameters $\delta=0.1$, $\varphi=0.3$, $\Gamma_0=1$, and $N=150$. The colors mark the tilted degree of the corresponding eigenstates. We are interested in the most subradiant state with the energy  marked as a circle in Fig.~\ref{InterfaceS}(a) and its probability distribution is shown in Fig.~\ref{InterfaceS}(b). Such a state is approximated by an anti-symmetric combination~\cite{Molmer2019} of the single-excitation interface state and the most subradiant  single-excitation state at the left part of the qubit array, as discussed in Appendix~\ref{A1}.
We also find a kind of cross-like interface state which is a mixture of a bound state and a self-localized state 
with the energy marked as a square, see Fig.~\ref{InterfaceS}(c). Such interface states are induced by the interplay between topology, interaction and non-Hermitian coupling. These two kinds of interface states have much smaller decay rates $(\Gamma<10^{-6}\Gamma_0)$ than that of the bound edge states $(\Gamma\sim 10^{-2}\Gamma_0)$ in Figs.~\ref{InterfaceS}(d, e). Importantly, the decay rate for the interface states decreases as the qubit number increases. However, for the bound edge states (marked as `$+$' and `$\times$') either localized at the left or right edges, the decay rates are in the order of $10^{-2}\Gamma_0$, which are in the same order of those in Fig.~\ref{TopoState} even with the increase of the system size.

\section{Summary and discussion}
\label{SecV}

We have studied topological properties of bound photons propagating in spatially-modulated atomic arrays coupled to a waveguide.
We have calculated the center-of-mass bands and characterized the topology of bound-state bands with the Chern numbers.
In contrast to many conventional systems realizing the bulk-edge correspondence, our system supports radiative bound-edge
states which can be associated with both the bandgaps and also bulk bands. 
When  the modulation phase is modified and and  passes through the critical points,
the edge states localized in the left part of the array transform to those localized in its right part, or vice versa.
When two arrays of qubits with different spatial modulations are connected, they can support long-lived interface topological states.

Previously, the breakdown of conventional bulk-edge correspondence has been discussed in either non-Hermitian or interacting systems.
For non-interacting non-Hermitian systems, the generalized bulk-edge correspondence has been suggested by employing non-Bloch topological invariants~\cite{yao2018edge} or bi-orthogonal polarization~\cite{Kunst2018}. For interacting Hermitian systems, the breakdown of bulk-edge correspondence may be explained by the virtual defect induced by particle-particle interactions. However, both non-Hermiticity and interaction coexist in our systems, making this type of problem much more complicated~\cite{Liberto2016}, so further study is required to uncover the principle of bulk-edge correspondence in our system.

Recently, topological features have been employed to improve the optical properties and control the optical response.
Out-of-plane-scattering losses, a hindrance for high-$Q$ resonators, can be suppressed by the topological nature of the bound state in continuum~\cite{jin2019topologically}. Our previous study has revealed that  double-excited subradiant states enable a longer time for light-matter interaction, and they can enhance the inelastic scattering of photon pairs~\cite{Ke2019}. An interesting avenue for future exploration of these kind of systems is to study whether the topological interface states can be used to manipulate the correlation and entanglement of photon pairs.

\acknowledgements{This work was supported by the Australian Research Council. C.L. was supported by the National Natural Science Foundation of China (NNSFC) (grants No. 11874434 and No. 11574405). Y.K. was partially supported by the Office of China Postdoctoral Council (grant No. 20180052), the National Natural Science Foundation of China (grant No. 11904419), and the Australian Research Council (DP200101168). A.V.P. acknowledges a support from the Russian Science Foundation (Project No. 19-72-00080). A.N.P. has been supported by the  Russian President Grant No. MD-243.2020.2.}

\appendix
\section{Approximation of the interface state}
\label{A1}

Here, we show how the interface state can be approximated by an antisymmetric combination of the single-excitation interface state, $|\psi_{int}^{(1)}\rangle$, and the most subradiant state,  $|\psi_{L}^{(1)}\rangle$ at the left part of the array.
By diagonalizing the Hamiltonian~\eqref{Ham} with $z_j$ arranged as Eq.~\eqref{Position} in the single-excitation subspace, we 
obtain the interface state $|\psi_{int}^{(1)}\rangle=\sum_j c_j|j\rangle$, with the longest lifetime.
By diagonalizing the Hamiltonian~\eqref{Ham} with $z_j=j + \delta \cos (2\pi j/3)$ $[j \le N/2]$ in the single-excitation subspace,
we can also obtain the most-subradiant single-excitation state, $|\psi_{L}^{(1)}\rangle=\sum_j f_j|j\rangle$ with $f_{j>N/2}=0$.
The ansatz of the most-subradiant interface state $|\psi_{int}^{(2)}\rangle=\sum_{j,l}\psi_{j,l}|j,l\rangle$ has the anti-symmetric form~\cite{Molmer2019}
\begin{eqnarray}
\psi_{j,l}=\begin{cases}
c_{j}f_{l}-f_{l}c_{j},&(j \ge l);\\
-(c_{j}f_{l}-f_{l}c_{j}),&(j \le l).
\label{ansatz}
\end{cases}
\end{eqnarray}

\begin{figure}[!b]
	\includegraphics[width=0.5\textwidth]{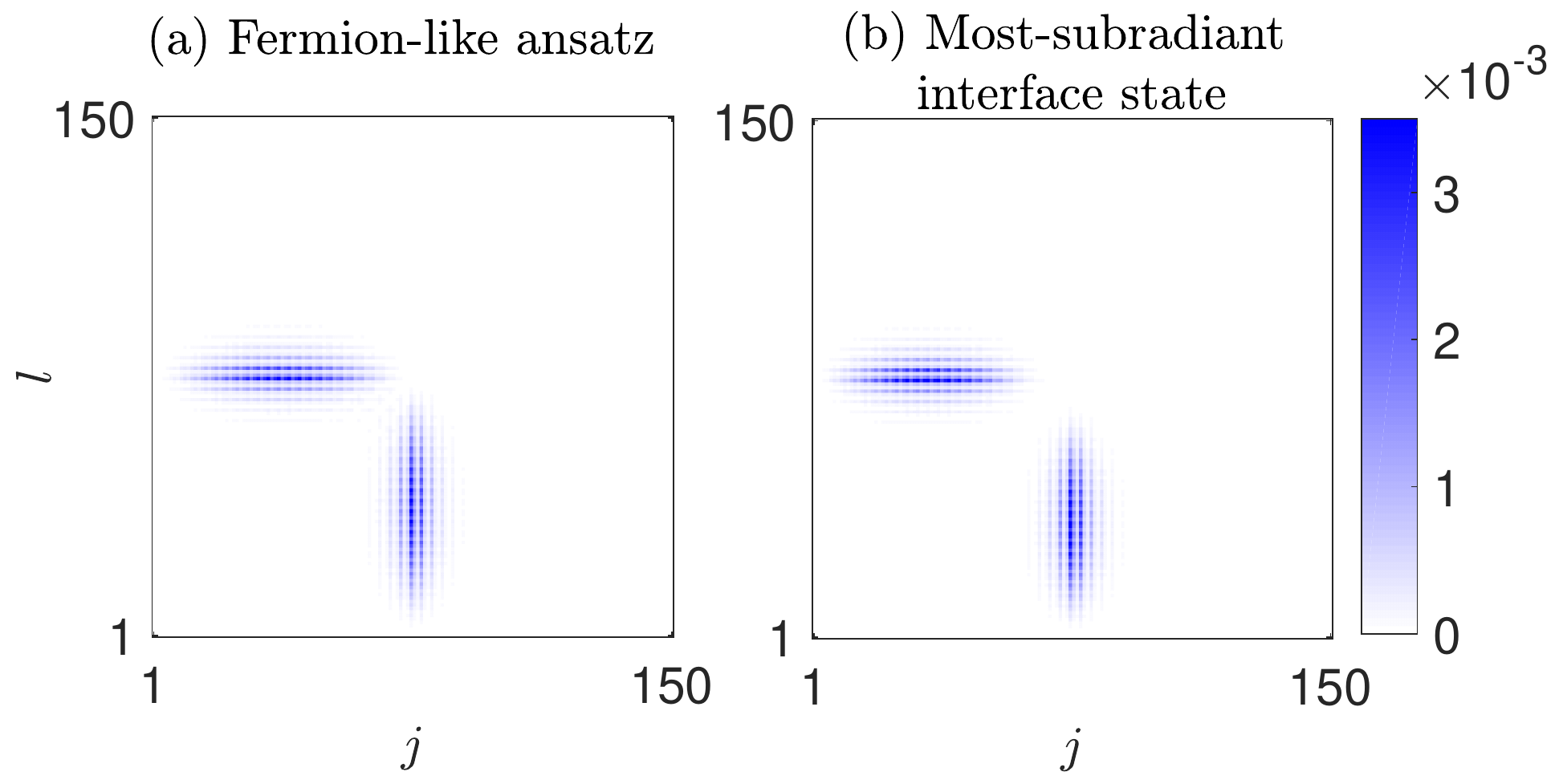}
	\caption{Probability distribution $|\psi_{j,l}|^2$ as functions of positions $j$ and $l$ for (a) the fermion-like ansatz and (b) the most-subradiant interface state. }\label{Supp}
\end{figure}

In Fig.~\ref{Supp}, we show the probability distributions $|\psi_{j,l}|^2$ as functions of the positions $j$ and $l$ for (a) fermion-like ansatz and (b) most-subradiant interface state. The parameters are chosen the same as those in Fig.~\ref{InterfaceS}.
It is clear that the fermion ansatz~\cite{Molmer2019} well describes the major features of the most-subradiant interface state of two excitations.

%\bibliography{2PhotoTopo}

%merlin.mbs apsrev4-1.bst 2010-07-25 4.21a (PWD, AO, DPC) hacked
%Control: key (0)
%Control: author (0) dotless jnrlst
%Control: editor formatted (1) identically to author
%Control: production of article title (0) allowed
%Control: page (1) range
%Control: year (0) verbatim
%Control: production of eprint (0) enabled
%

\end{document}